\documentclass[10pt,twocolumn]{article}
\usepackage{times}
\usepackage[a4paper,margin=0.75in]{geometry}
\usepackage{times}  
\usepackage{helvet}  
\usepackage{courier}  
\usepackage[hyphens]{url}  
\usepackage{graphicx} 
\usepackage{natbib}  
\usepackage{caption} 
\usepackage[usenames,dvipsnames]{xcolor}

\newcommand{\out}[1]{}

\title{A Datalake for Data-driven Social Science Research}

\author{Puneet Arya$^1$\\puneet.arya@ashoka.edu.in \and Ojas Sahasrabudhe$^1$ \and Adwaiya Srivastav$^1$ \and Pratha Pratim Das$^1$\\ppd@ashoka.edu.in \and Maya Ramanath$^2$\footnote{Work done while visiting Ashoka University.}\\ramanath@cse.iitd.ac.in}
\date{$^1$ Ashoka University, Sonipat, India\\
$^2$ IIT Delhi, New Delhi, India
}

\begin{document}

\maketitle

\begin{abstract}
Social science research increasingly demands data-driven insights, yet researchers often face barriers such as lack of technical expertise, inconsistent data formats, and limited access to reliable datasets. In this paper, we present a Datalake infrastructure tailored to the needs of interdisciplinary social science research. Our system supports ingestion and integration of diverse data types, automatic provenance and version tracking, role-based access control, and built-in tools for visualization and analysis. We demonstrate the utility of our Datalake using real-world use cases spanning governance, health, and education. A detailed walkthrough of one such use case—analyzing the relationship between income, education, and infant mortality—shows how our platform streamlines the research process while maintaining transparency and reproducibility. We argue that such infrastructure can democratize access to advanced data science practices, especially for NGOs, students, and grassroots organizations. The Datalake continues to evolve with plans to support ML pipelines, mobile access, and citizen data feedback mechanisms. 
\end{abstract}

\maketitle

\section{Introduction}
\label{sec:intro}

There is a continuing rise in demand for data-driven methods in social science research, especially in emerging economies. While data science methods are already extensively used in disciplines such as finance, genomics, engineering, etc., social science research is only now looking to fully exploit these tools. The former disciplines have long benefited from high-quality datasets and advanced IT infrastructure, driven as they are by both government as well as corporate funding. In contrast, the social science sector suffers from a number of factors, the chief of which are the lack of availability of reliable and standardised datasets (especially those datasets that are collected by individual groups) and the lack of skilled data professionals who can work with social sector data. Further, the questions addressed by social science research are vastly interdisciplinary, drawing from diverse domains such as health, education, governance, urban planning, etc. (Figure \ref{tab:usecases} shows a few examples). Therefore, basic data science practices such as data reuse and reproducibility of results are difficult because of the lack of reliable datasets, which adds to the burden of researchers in these areas. A possible solution is create an eco-system of datasets and researchers through a easy-to-use Datalake infrastructure that strictly follows standard data governance principles, thus increasing trust in the analysis and outcomes of research.

We talked to many social scientists to understand these issues in detail and evolved a set of goals for our Datalake infrastructure tailored specifically to enable interdisciplinary research. Our Datalake design focuses on the following goals:
\begin{itemize}
    \item \emph{Provenance and version control} for different dataset types such as tables, text, and images. These are two very important principles of data governance that leads to increased trust in datasets and encourages data reuse. The provenance of a dataset refers to how exactly the dataset was derived (or curated) from raw sources. Knowing the provenance allows users to recreate the dataset if needed. Similarly, version control ensures that users have access to all versions of a dataset and are able to trace exactly how the dataset has evolved into the current version.
    
    \item \emph{Provide an integrated environment} for \emph{analyses and visualisations}. This helps users maintain their datasets as well as  analyses in the Datalake itself without having to manually transfer files back and forth to different machines for different analytical tasks. All data items (intermediate results, plots, etc.) generated within the Datalake can be saved by the user as needed, and their provenance tracked automatically by the system.
    
    \item Enable users to easily \emph{share and search for datasets}. This feature is key to building a good eco-system of data. Data sharing happens not only by making a dataset public, but also by linking to the dataset in research papers or other articles. The search feature should not only allow users to search for data items within the Datalake, but also allow them to search standard, external data sources for relevant datasets, download and save them into the Datalake.

\end{itemize}

\subsection{Use Cases}

\begin{table*}[ht]
    \centering
    \begin{tabular}{l|p{0.3\linewidth}|p{0.3\linewidth}|p{0.3\linewidth}}
    \hline
     \#&{\bf Task}    & {\bf Dataset Sources} & {\bf Features Required} \\
     \hline
      1& Analyze whether voter turnout is affected by infrastructure access and literacy rates.   & Political Science + Urban Studies + Education &  Merge (at least) 3 different datasets based on their area/region, generate plots and map overlays.\\
     \hline
      2 & Analyze whether constituencies with women legislators correlate with female literacy and female school enrolment rates. & Political Science + Education & Ability to merge (at least) 2 different datasets based on their area/region, perform regression analysis, generate plots.\\
     \hline
      3&  Analyze the time required to restore drinking water connections in the aftermath of cyclonic storms. &  Disaster management + Urban Studies + Public Policy & Time-based merge of datasets, time-based plots, and map overlays.\\
     \hline
      4& Analyze whether income level and educational attainment have a meaningful correlation with infant mortality rate. & Economics  + Education + Health & Ability to merge multiple datasets based on IDs, and generate correlation plots.\\
      \hline
     5& Analyze whether greater digital access positively influences women's employment and use of modern family planning methods. & Digital Infra + Economics + Health & Ability to merge multiple datasets based on Time Period, Geographic location, and generate various charts.\\
     \hline
    \end{tabular}
    \caption{This table outlines basic analysis tasks, likely social science disciplines that collect and curate datasets relevant to the task, and features required from our Datalake to perform the analysis task.}
    \label{tab:usecases}
\end{table*}
\begin{figure}[ht]
    \centering
    \includegraphics[width=1.0\columnwidth]{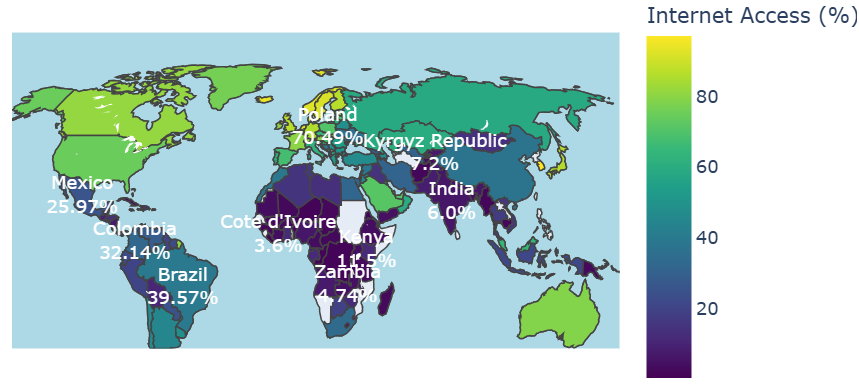}
\caption{\small \textbf{Data Visualization: Household Internet Access}}
\label{fig:geo-map-example-1}
\end{figure}
\begin{figure}[ht]
    \centering
    \includegraphics[width=1.0\columnwidth]{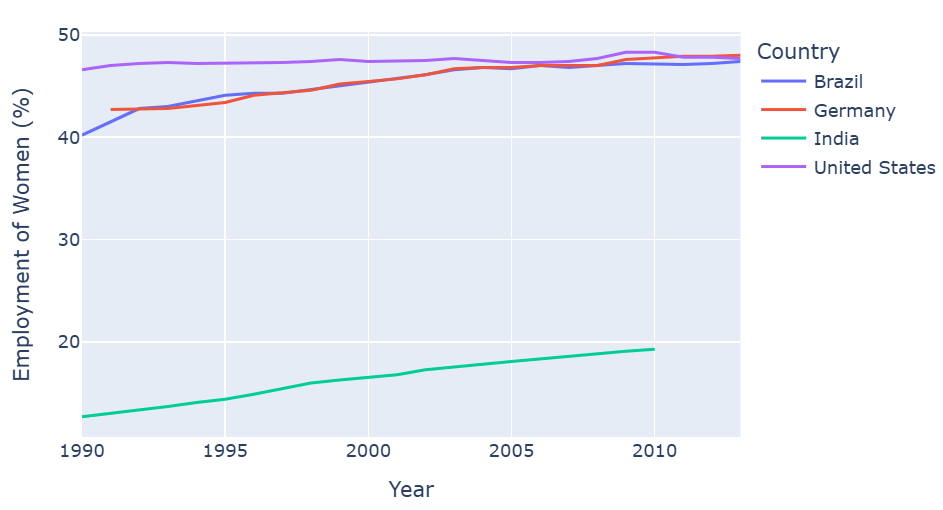}
\caption{\small \textbf{Data Visualization: Trends in Women Employment in non-agriculture sector across countries}}
\label{fig:line-chart-example-1}
\end{figure}
With this motivation, we now list a number of use cases that benefit from our Datalake infrastructure. Table \ref{tab:usecases} lists a few useful analysis tasks that are \emph{multi-disciplinary}. In particular, they require datasets that are typically collected and curated by social scientists with different kinds of expertise. Our Datalake was built precisely for easy sharing of such datasets and to enable analyses of the kinds of tasks listed in Table \ref{tab:usecases}. As an example, we show a set of visualisations generated for Task 5 from Table \ref{tab:usecases} in Figure \ref{fig:geo-map-example-1} and \ref{fig:line-chart-example-1}. These are potentially intermediate visualisations that help understand the various datasets. The first choropleth map shows the percentage of household internet access across the world. The second line chart shows the percentage of women employed in non-agriculture sectors in countries from 1990 till 2013.

\subsubsection*{Related Work} While many commercial, open source and academic/community oriented Datalake offerings exist \cite{s3,apache-hudi,apache-iceberg,azure,biglake,harvarddataverse, zotero, mendeley}, none of them are geared for interdisciplinary researchers in an up-and-coming area such as social science research\footnote{Please note that our comments on the status of social science research are especially true for developing countries.}.  In particular, while many of these systems address issues of scalability, security, and analysis tools, none address our specific goals of enabling trust and reproducibility through in-built support for provenance tracking and versioning along with dataset-type specific analysis tools.

\subsubsection*{Contributions} Our main contribution is the development of a Datalake infrastructure to enable inter-disciplinary social science research. We describe the components of our Datalake along with key governance principles (provenance and version control) that enable the creation of an eco-system of datasets that are trustworthy and reusable. We show through examples, the utility of our Datalake.

\subsubsection*{Organization} The rest of this paper is organized as follows. In Section \ref{sec:lifecycle}, we give an overview of the Data Lifecycle which is typical of data-driven research. In Section \ref{sec:datalake} we highlight the key features of the Datalake that support the data lifecycle, while adhering to the goals highlighted earlier in this Section. In Section \ref{sec:usecase}, we give a walkthrough of a concrete analytical task that benefits from the use of our Datalake. Finally, we conclude in Section \ref{sec:conc}.

\section{The Data Lifecycle}
\label{sec:lifecycle}

\begin{figure}[ht]
    \centering
    \includegraphics[width=\columnwidth]{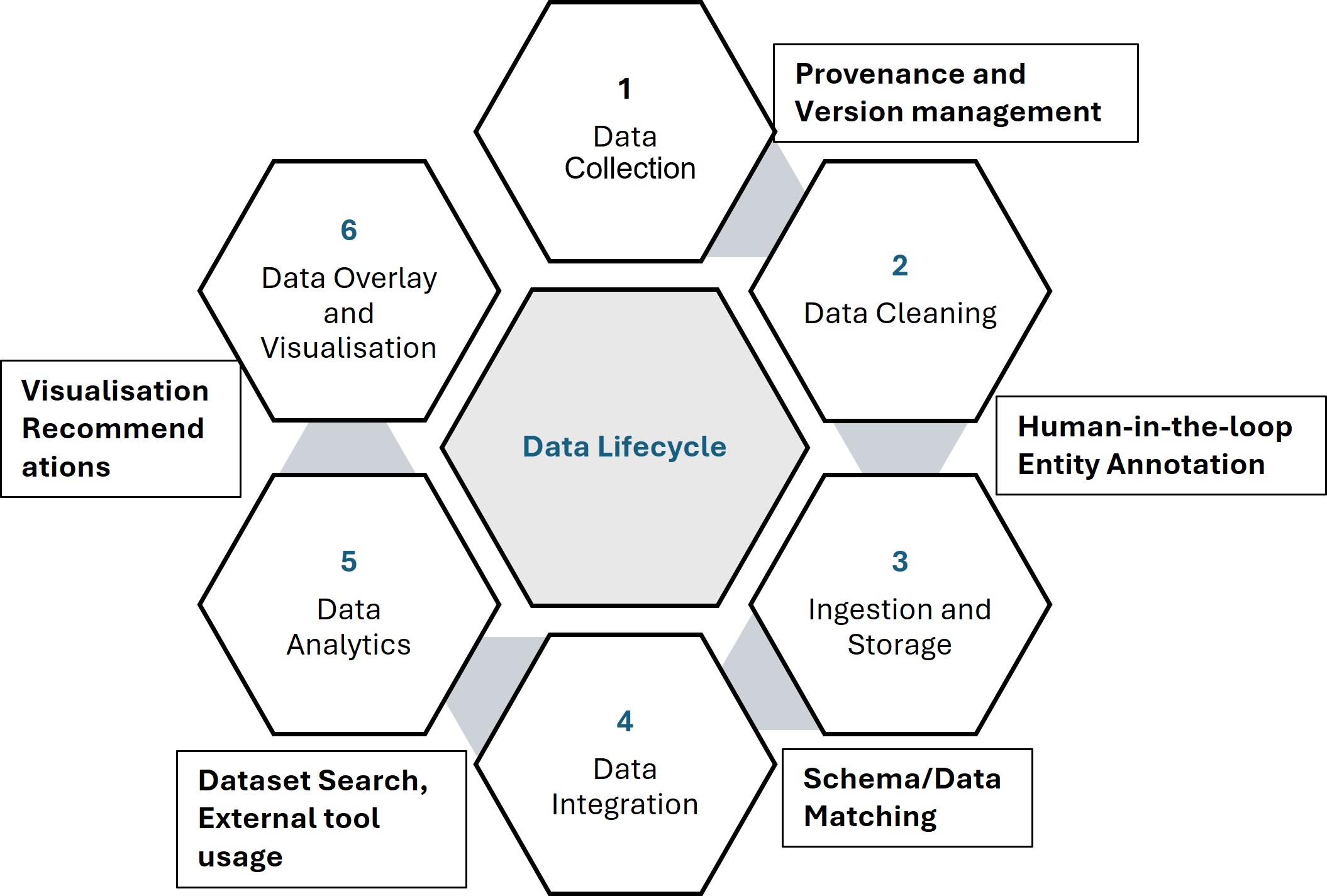}
    \caption{\small The typical lifecycle of datasets. The boxes highlight technical challenges associated with supporting the various stages of the lifecycle.}
    \label{fig:data-lifecycle}
\end{figure}

\begin{figure}[ht]
    \centering
    \includegraphics[width=1.0\columnwidth]{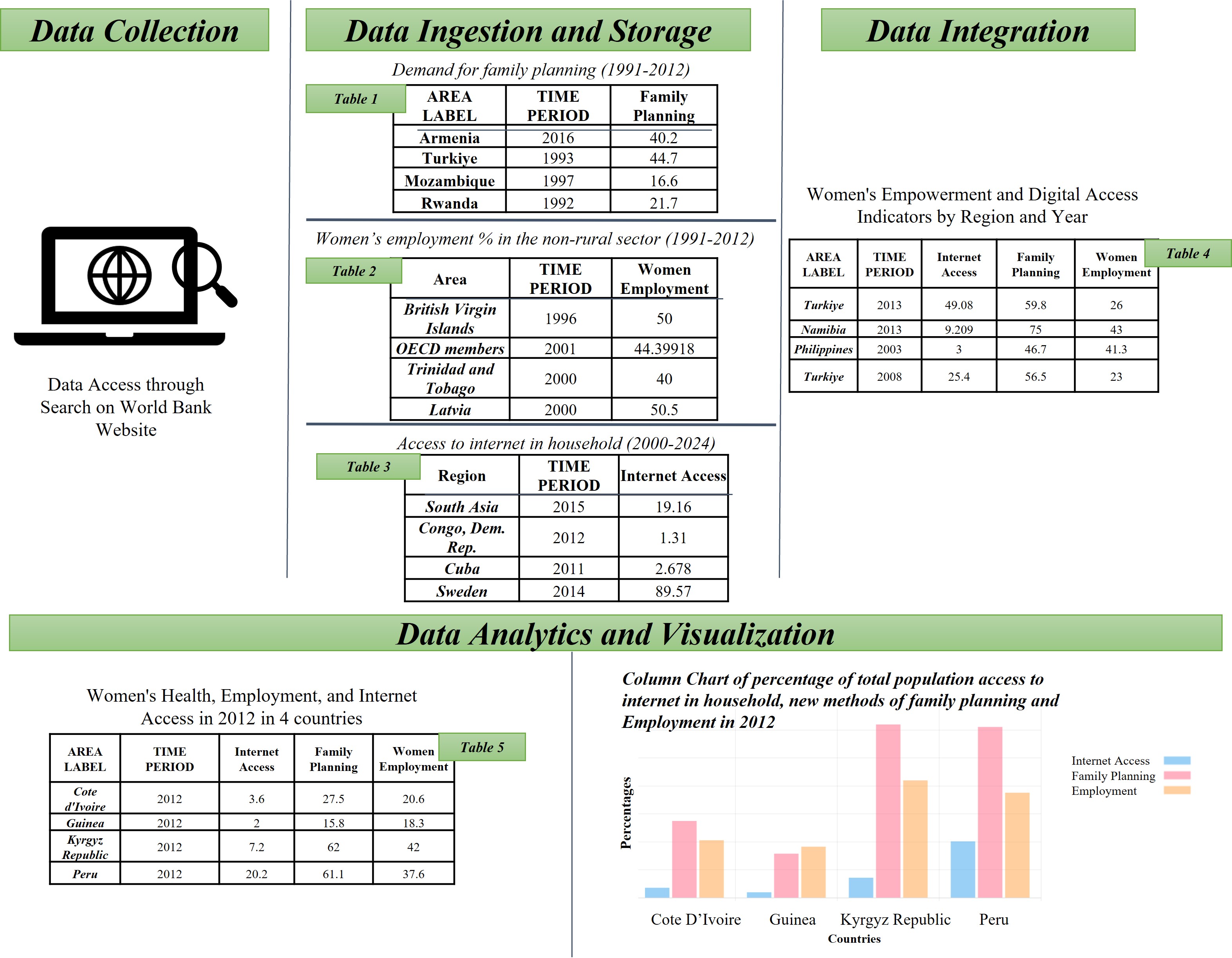}
\caption{\small \textbf{Data Lifecycle: Family Planning, Employment and Digital Access}: Tables 1, 2, 3 are the ingested datasets. Table 4 results from a merge of Tables 1, 2 and 3. Table 5 is a result of selecting specific rows from Table 4 for analysis. The bar chart is derived from Table 5.}
\label{fig:data-lifecycle-example-1}
\end{figure}

Figure \ref{fig:data-lifecycle} gives an overview of the data lifecycle showing the various kinds of processing a dataset may undergo: starting from Data Collection to analysis and insights in the form of Visualisations. Based on the insights gained, the whole process may restart with new data. In order to support this data lifecycle, it is critical that the processed data at each step is \emph{reliable}, \emph{accessible}, and \emph{reproducible}. Our Datalake aims to achieve these goals for \emph{each dataset}.

We describe this lifecycle through an example -- the last usecase (\#5) in Table \ref{tab:usecases}. Figure \ref{fig:data-lifecycle-example-1} illustrates several of the components in the data lifecycle for this task. Briefly, this task involves looking at datasets for multiple countries across the world and studying the relationship between access to family planning, internet at home, and employment opportunities for women. The user may go through the following steps:

\begin{itemize}
    \item {\bf Data Collection}: The user searches the website of the World Bank (and potentially other websites) and  collects datasets related to family planning, internet access, and employment numbers.

    \item {\bf Data Cleaning}: 
    Data cleaning and pre-processing are essential steps in any data analysis workflow. Even datasets obtained from reputable sources may require further processing to ensure consistency, address missing values, handle outliers, or adapt the data for specific research objectives. 

    \item {\bf Ingestion and Storage}: The first step in studying these datasets is to store them in a suitable tabular storage format (such as in a Database System). This enables the user to issue queries and perform the next step, integrating the collected data into a form that is helpful for visualisation and analysis.

    \item {\bf Data Integration}: Data integration involves two steps in this example: first, among the three tables, the column names have to be reconciled. That is, ``AREA LABEL" in Table 1, ``Area" in Table 2, and ``Region" in Table 3, all refer to the same type of entity. Second, once this uniformity is established, the user can now perform a 3-way join with Table 1, Table 2, and Table 3, which gives her an integrated view of the three variables she is studying, as illustrated in Table 4.

    \item {\bf Data Analytics}: This step may involve different kinds of analyses, including regression analysis, or more complicated machine learning models. However, in our example, we show a very simple analysis -- that of isolating the numbers for only the year 2012. This corresponds to a "Select" query over Table 4, derived in the previous step.

    \item{\bf Visualisation}: Note that visualizations can be generated for any table at any intermediate step, as shown in Figures \ref{fig:geo-map-example-1} and \ref{fig:line-chart-example-1}. In Figure \ref{fig:data-lifecycle-example-1}, we simply show a column chart based on Table 5 generated in the previous step. However, it is possible to do more complex visualisations, including choropleth maps.
\end{itemize}

We now describe the features of our Datalake that support each of the steps highlighted above and describe additional possibilities for more complex data processing and analysis. Further, each step of the data lifecycle is tracked and each intermediate step is made reproducible through provenance tracking.

\section{Datalake model and features}
\label{sec:datalake}

\begin{figure}[ht]
    \centering
    \includegraphics[width=\columnwidth]{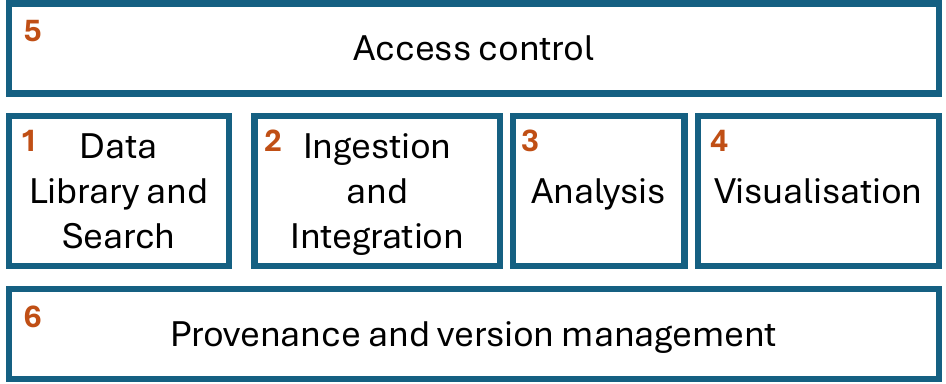}
    \caption{\small Components of the Datalake}
    \label{fig:datalake}
\end{figure}

Our Datalake architecture is geared towards managing the data lifecycle described in Section \ref{sec:lifecycle} and aims to fulfill the goals outlined in the Introduction. Currently, the Datalake supports tabular data, images, pdfs, and entire databases -- users can upload, search, and analyze any of these dataset types.

Figure \ref{fig:datalake} depicts the main components of our Datalake. We describe these components below and highlight how they relate to the data lifecycle and the example of the previous Section.

\paragraph{Data Search and Exploration:} Analysis typically starts by collecting data that is relevant to the task. The Datalake supports this task by enabling \emph{online discovery} of datasets by providing two kinds of search functionality (block \#1 in Figure \ref{fig:datalake}.

First, \emph{internal search} allows users to search all \emph{public} data items\footnote{We use the term data item to indicate an uploaded dataset (table, image, pdf, etc.) with a name and description provided by the user.} using keywords and save them in a separate \emph{Workspace} for later analysis. The search process is currently supported by a simple keyword-based search engine that indexes all metadata (that is, dataset name and description, user who uploaded the dataset, time of upload, etc.) as well as the content. A \emph{multi-modal} retrieval system that supports text, image,  and tabular search in an integrated manner is currently under development.

\begin{figure}[ht]
    \centering
    \includegraphics[width=1.0\columnwidth]{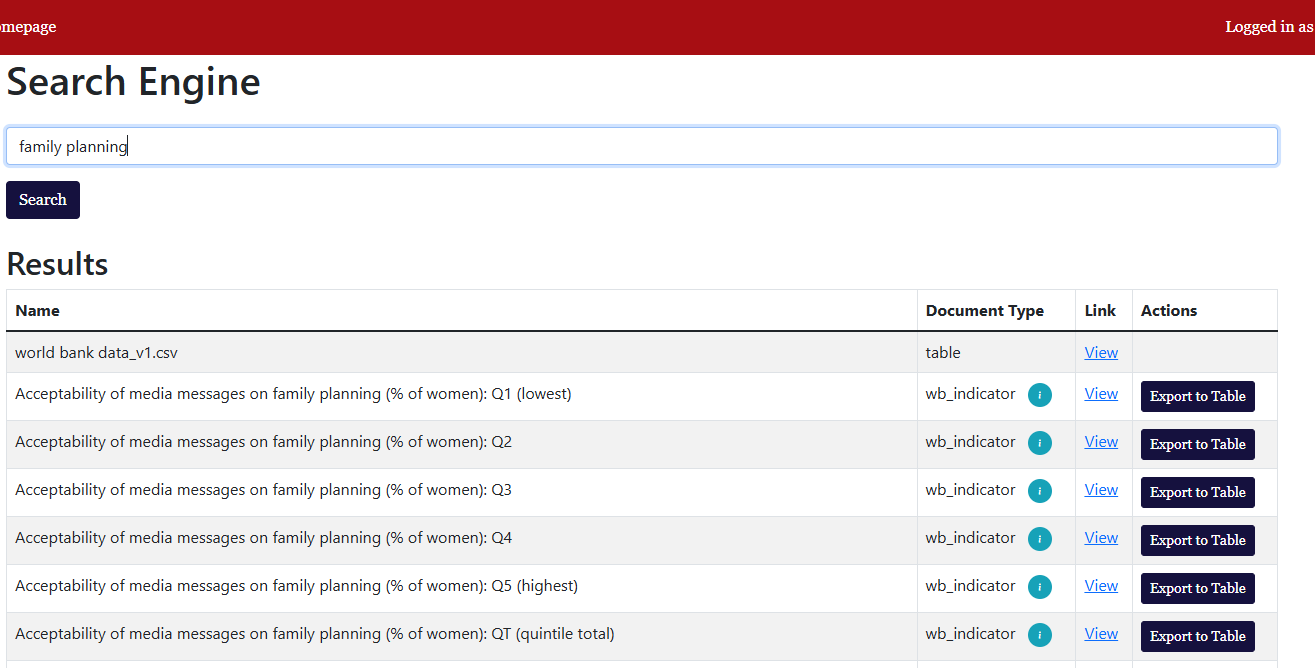}
\caption{\small \textbf{Data Collection:} This screenshot from our Datalake shows a search "family planning" in the Data Library and results from the World Bank website.}
\label{fig:data-collection-example-1}
\end{figure}

Second, a \emph{data library search} allows users to search certain popular websites (such as the World Bank website) from \emph{within} the Datalake. The Datalake provides a separate search page with a list of website APIs supported. The user fills in relevant keywords into a single search box, and the system ensures that each selected website is automatically searched and the results are displayed to the user. The user can then choose one or more of these datasets to ingest and store in her Workspace for later analysis. The data library reduces user effort by eliminating the need to search multiple websites separately, temporarily store interesting data sets, and then upload them to the Datalake. Figure \ref{fig:data-collection-example-1} shows a screenshot from our Datalake illustrating how a user can simply search the World Bank website from within the Datalake and easily view or import datasets.

\paragraph*{Ingestion and Integration:} As the user uploads (or imports through a data library search) her dataset into the datalake, the ingestion and integration module (block \#2 in Figure \ref{fig:datalake}) performs a set of pre-processing steps. Depending on the dataset type, specific processing steps are applied as follows.

\noindent
\emph{Tabular data and Databases module}: Tabular data is typically ingested through csv files or Excel sheets. This module stores relevant \emph{user sourced} metadata like column types (Numerical, Categorical, Location, etc.) and creates appropriate tables in a database; this ensures that the various analysis tools, such as filter, merge, etc., are easily implemented, and recommending visualisations becomes easier.

In order to make the process of identifying column types easier, the module queries an LLM to determine the type of each column from a pre-specified list of column types. The LLM is prompted with a list of all columns with a few example rows, and asked to identify the type of the column as either `Categorical', `Numerical', `Location' (including lat/long pairs and location names), `URL', `HTML/Text', or `Temporal'. The user is then left with the task of approving the type inferred by the system or correcting it by selecting a different type from a drop-down menu. For instance, in Figure \ref{fig:data-lifecycle-example-1}, columns AREA LABEL and TIME PERIOD were detected as Location and  Time, respectively, for Table 1 (similarly for the corresponding columns of Tables 2 and 3) without requiring a correction from the user.

A similar pre-processing is done for database uploads as well -- the entire database schema is created along with constraints. However, no functions, custom types etc., are allowed for security reasons.

All content from these tables, along with user given meta-data (including the name of the dataset and a description) are indexed to enable search.

\noindent
\emph{Image and Text module} Text is mainly extracted from PDF files. The PDF files may be scanned or well-formed. Each PDF file that is uploaded is sent through a processing pipeline. Apart from text extraction, the PDF file is broken down into multiple pages. Images and tables are also extracted and stored separately. All content, including the meta-data provided by the user is indexed for search.

Images do not undergo any specific pre-processing steps. They are simply ingested, stored and indexed for search.

\paragraph*{Analysis:} 
The analysis tools provided by the Datalake are of two kinds to enable both simple and complex analysis (block \#3 in Figure \ref{fig:datalake}.

\noindent
\emph{Simple analysis tools} are dataset-type-specific. First, Tabular data items and  Database tables can be \emph{filtered} by row or columns, or \emph{merged ("joined")} with two or more tabular data items (these data items, in turn could be discovered through an internal search). The resulting data item can then be stored as a new data item with its own name and description, and visualised using the Datalake's Visualisation features. As we describe later, these analysis operations are carefully tracked and their provenance preserved for reproducibility.

Second, as mentioned earlier, PDFs are automatically broken down into multiple pages and images. These pages and image items from the same or different PDFs can be combined into new data items with their own names and descriptions.

\noindent
\emph{External analysis tools} While these simple tools help with preliminary analysis (in fact, almost all of our examples tasks from Table \ref{tab:usecases} could be undertaken with only these simple tools), more sophisticated frameworks are required for advanced studies. Therefore, we integrate \emph{external tools} such as R within the Datalake framework. Users can open the R environment from within the Datalake. The data items available in their own Workspace is automatically shared in a common disk space accessible by both R as well as the Datalake. Users can perform analysis, generate new data and charts, and save these new data items seamlessly back in the Datalake through the common disk space.

\paragraph*{Visualizations:}
Our Datalake offers a number of \emph{standard visualisations} such as bar/line/scatter plots, 2D and 3D visualisations (block \#4 in Figure \ref{fig:datalake}. Examples of these visualizations are shown in Figures \ref{fig:geo-map-example-1} (map overlay), \ref{fig:line-chart-example-1} (line chart) and \ref{fig:data-lifecycle-example-1} (bar chart). Note that the visualisations in Figures \ref{fig:geo-map-example-1} and \ref{fig:line-chart-example-1} are from individual tables where the user may be trying to explore the datasets, the bar chart in Figure \ref{fig:data-lifecycle-example-1} shows an integrated view of the analysis performed (Table 5 in Figure \ref{fig:data-lifecycle-example-1}).

Apart from the visualization tools available within the Datalake, users can use external tools (as described before) for their analysis and plot generation. These new data items can be seamlessly ingested into the Datalake as images and made searchable.


\subsection{Data Governance}
\label{sec:governance}
While all the features we have described so far help social scientists do analysis and share data in a user-friendly way, a key goal of the Datalake is to develop an ecosystem of data where users can trust the data items available and reuse them in their own research. In order to fulfil this goal, our Datalake provides a principled way of tracking the evolution of datasets through two inter-related mechanisms: provenance and version control. Further, the Datalake provides rigorous access control to ensure that only a set of approved members have permissions to upload, update, and share data items. We describe these components in more detail below.

\paragraph*{Access control} Users can store their data items in workspaces and sub-workspaces (similar to a directory structure).  Three levels of access permissions -- admin, staff, and guest -- determine exactly which workspaces and data items a user has access to and what operations a staff member is allowed to perform in those workspaces. Once a user is designated as an admin, the user can create their own workspaces and add staff members to each workspace. Further, each staff member may be given permissions only for specific actions within the workspace. For example, the admin of a certain workspace may decide that only certain staff members are allowed to upload data, but all staff members are allowed to create and store visualizations. The ability to set such fine-grained permissions will enable users to be in control of their data items.

In addition to the permissions regime, each data item can be designated "public" or "private" by the admin. If a data item is private, only members of the workspace to which the data item belongs are able to view and manipulate it. A private data item is not returned as a search result for non-members of that workspace. In contrast, a public data item is viewable by all. These data items can be searched, used for analyses, and used for visualizations. Further, the data item or workspace can be linked through a URL that points exclusively to it. This enables users to link to their datasets in articles or research papers.

\paragraph{Provenance and version management} 
A robust provenance and version management system is central to the Datalake’s design, ensuring transparency, trust, and reproducibility throughout the data lifecycle. As described before, the data lifecycle implies that datasets undergo multiple transformations -- cleaning, integration, enrichment, etc. -- and it is essential for researchers to trace each step in this journey. With reliable provenance, data driven conclusions remain accountable and reproducible. 

Our Datalake addresses this challenge by embedding provenance tracking at every stage of the data lifecycle, starting from the ingestion of a dataset into the Datalake to integration, transformation, and analysis. We follow W3C recommendations for provenance management \cite{prov, prov-w3c}. Each data item is accompanied by a structured provenance record capturing its source, all applied operations, and the user or tool responsible for each step. This record is available for viewing to all with access to the data item, ensuring transparency. Users can use the provenance information in at least two ways: i) verify how the data item has been produced. This is important, for example, when plots or charts are reported in research papers, and readers need to verify the claims made with those charts, ii) redo the same steps for a different data item(s) to compare and contrast a previous result.

\begin{figure}[ht]
    \centering
    \includegraphics[width=\columnwidth]{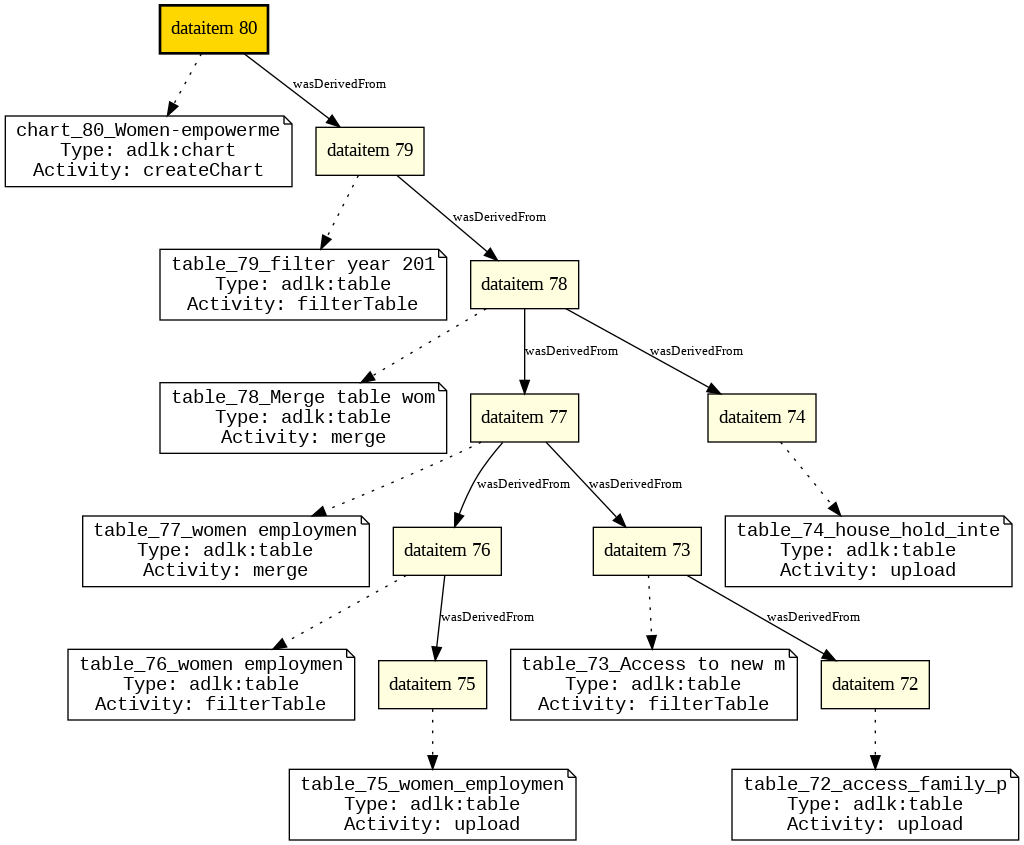}
\caption{\small \textbf{Column Chart Provenance}: i) `dataitem 80' at the root of the tree is the Column Chart. ii) Solid lines show the lineage of each data item, dotted lines point to the data item's attributes. iii) The white boxes show the attributes of the data item including the type, name of the data item and the operation used to create it. Other details, such as parameters of the operation (filter value, merge key, chart axes, etc,) are internally recorded. }
\label{fig:prov-example-1}
\end{figure}

The provenance module offers a \textbf{visual provenance tree} that graphically shows the \emph{lineage} of each data item. Researchers can easily trace back to original sources, understand intermediate operations, and study how a dataset evolved over time. An example provenance graph for the column chart in Figure \ref{fig:data-lifecycle-example-1} is shown in Figure \ref{fig:prov-example-1}.


Closely related to the provenance tracking is the version management. The latter is particularly important when multiple users collaborate on the same dataset. For example, an initial version of a data item may be augmented with additional rows by a different user. It is then important to keep track of this new version so that the evolution of the same data item is tracked. Further, data items such as charts that have been created for a data item will change if a new version of the same data item is uploaded. Our Datalake ensures that such changes are automatically handled -- that is, a new version of the chart is automatically generated and its provenance stored. In general, whenever a new version of a data item is made available, all data items derived from it also need to be recreated and saved as new versions. The provenance tracking helps us do precisely that.

\section{Use Case Demonstration}
\label{sec:usecase}

\begin{figure*}[ht]
    \centering
    \includegraphics[width=\linewidth]{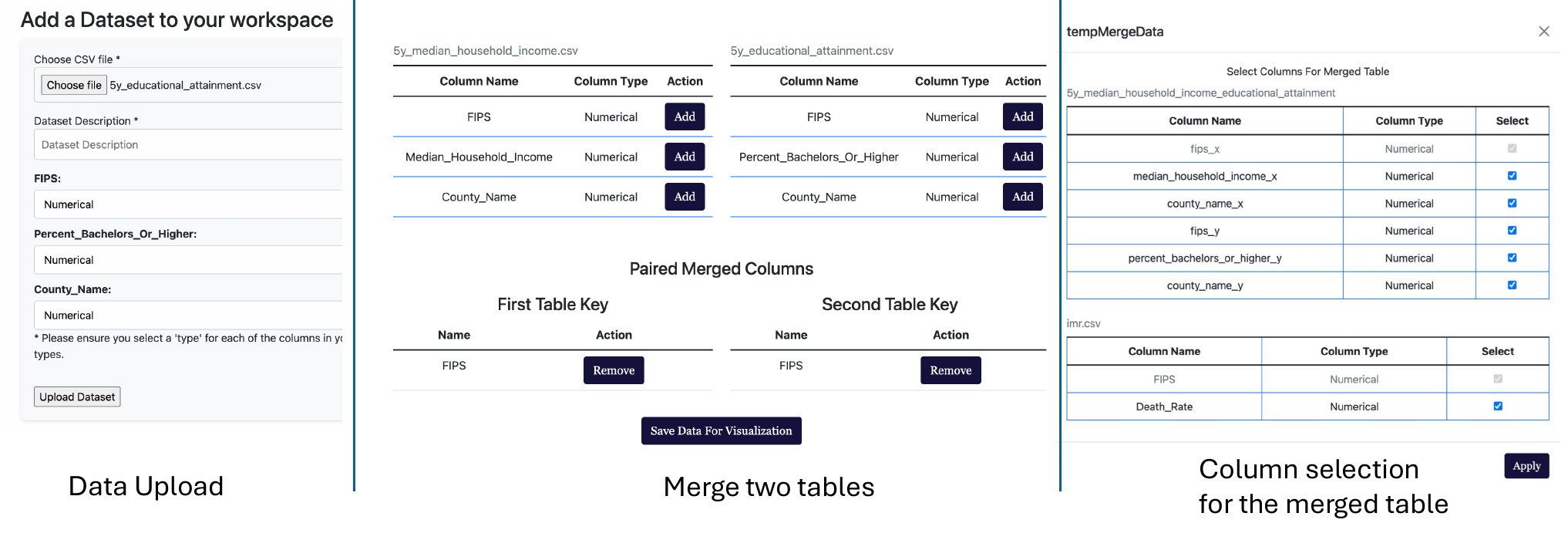}
    \caption{\small \textbf{Screenshots from the Datalake: Data Ingestion and Analysis:} i) The Data Upload shows automatic inference of column types, ii) Merge (join) tables feature automatically infers the join columns, with a facility to add or remove columns as join attributes, iii) Column Selection allows the user to select which columns should be in the output of the merged (joined) table.}
    \label{fig:ingestion_workflow}
\end{figure*}

In Sections \ref{sec:lifecycle} and \ref{sec:datalake} we gave an overview of the data lifecycle and the features in our datalake that help in implementing this lifecycle. In this Section, we give a concrete end-to-end application of the Datalake. Our task is to comprehensively analyze the correlation between median household income, educational attainment, and infant mortality rate (IMR) across United States counties. This walkthrough illustrates how our Datalake supports each stage of the data lifecycle, from initial data generation to final insights.

\subsection{Data Collection and Cleaning}
A researcher first needs to acquire data from at least two sources. In our case: i) \textbf{Income and Education Data:} From the U.S. Census Bureau's American Community Survey (ACS) \citep{USCensus2020}, ii) \textbf{Health Data:} From the CDC WONDER database \citep{CDCWonder}. This data is exported as a fixed-width text file. 

An easy way to do this is to make use of the data library of the Datalake, or alternatively, to go directly to these websites to search for suitable datasets.

Following the data lifecycle in Figure \ref{fig:data-lifecycle}, we perform essential data cleaning steps before ingestion of these datasets into the Datalake\footnote{Note these steps are performed outside the Datalake at the moment.}: i) \textbf{Health Data Processing:} The CDC WONDER export contains header metadata, footer notes, and non-numeric annotations like `(Unreliable)' that must be removed, ii) \textbf{Format Standardization:} Convert the fixed-width text file to CSV format for compatibility with our Datalake ingestion system, iii) \textbf{Column Alignment:} Ensure consistent naming conventions across datasets (e.g., standardizing FIPS code column names).

This cleaning phase ensures data quality and compatibility before proceeding to the ingestion stage.

\subsection{Ingestion and Integration}
With cleaned data ready, we proceed to ingest the datasets into our Datalake. We create a new Workspace named `US Counties' and upload our datasets with appropriate titles and descriptions (this meta-data and the contents of the table are automatically indexed for search). Figure \ref{fig:ingestion_workflow} shows a screenshot of the process.

The three source datasets must then be combined into a single, analysis-ready table. The common link between them is the 5-digit county FIPS code. Within the Datalake, we perform a series of two `merge' operations using a simple interface (please refer to Figure \ref{fig:ingestion_workflow}: i) Join the Income and Education tables on the `FIPS' key, ii)  Join the resulting table with the Health (IMR) table on the `fips\_x' and `FIPS' key.

\subsection{Visualization and Analysis}

\begin{figure}[ht]
    \centering
    \includegraphics[width=\columnwidth]{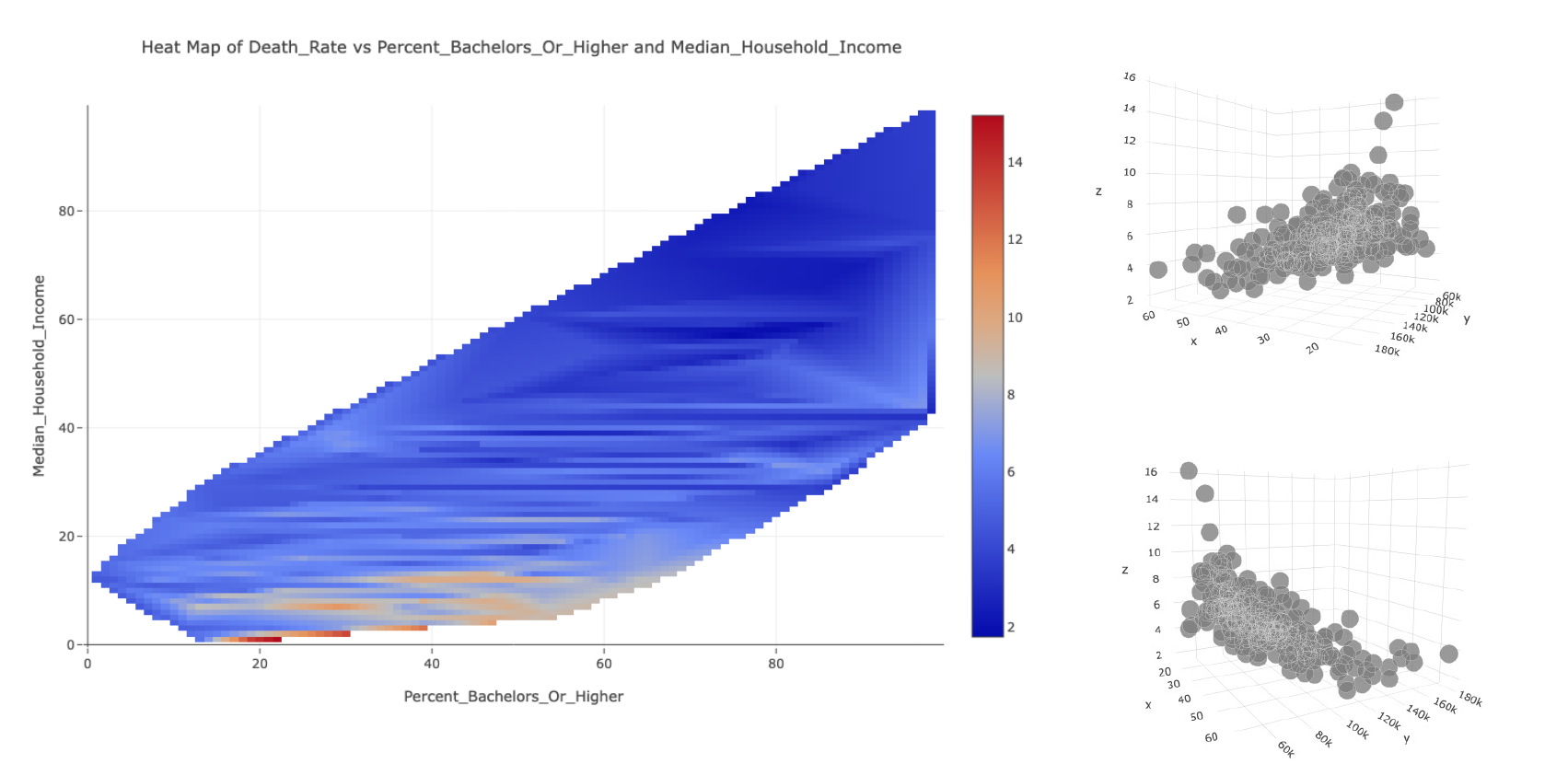}
    \caption{\small Visualizations created from the merged table depicted in Figure \ref{fig:ingestion_workflow}: \textbf{1.  Correlation Heat Map (left)}: A 2D heat map correlating Educational Attainment (x-axis) and Median Household Income (y-axis) with the average Infant Mortality Rate (the color). The scale ranges from blue (low IMR) to red (high IMR). \textbf{2. 3D scatter plot} (right top and bottom): 2 views of a 3D chart showing the relationship between MHI (x-axis), EA (y-axis), and IMR (z-axis)}
    \label{fig:usecase-visualization}
\end{figure}
With the integrated table ready, we can now generate visualizations directly within the Datalake without writing any code. Figure \ref{fig:usecase-visualization} shows two kinds of visualizations available within the Datalake. First, to understand the correlation between the variables, we create a \textbf{2D Heat Map} (left part of Figure \ref{fig:usecase-visualization}) with the following configuration: i) \textbf{X-Axis:} `Percent\_Bachelors\_Or\_Higher', ii) \textbf{Y-Axis:} `Median\_Household\_Income', iii) \textbf{Parameter (Color):} `Death\_Rate', iv) \textbf{Interpolation Method:} Linear.

Second, to gain deeper insights into the relationships between income, education, and infant mortality, we create a \textbf{3D scatter plot} (right part, top and bottom, of Figure \ref{fig:usecase-visualization}) with the following configuration: \textbf{X-Axis:} `Median\_Household\_Income', \textbf{Y-Axis:} `Percent\_Bachelors\_Or\_Higher', \textbf{Z-Axis:} `Death\_Rate'.



The visualizations provide immediate insights. The heat map in Figure 
\ref{fig:usecase-visualization} grows a darker shade of blue as median household income increases and has yellow/orange streaks where median household income is low.

The chart is bluish throughout the x-axis, suggesting that educational attainment is not as strongly correlated with Infant Mortality Rate as Median Household income. Despite that, the chart attains its darkest shade when both x and y axes are over 60.

The 3D scatter plot in Figure \ref{fig:usecase-visualization} reinforces these findings by showing clear spatial separation between high and low socioeconomic counties in three-dimensional space. One may notice in the heatmap that at median household income levels below 10k and education attainment below 20\%, there are again dark streaks of blue. This is because counties with less than a certain number of deaths suppress their data.


\subsection{Provenance}

\begin{figure}[ht]
    \centering
\includegraphics[width=0.9\columnwidth]{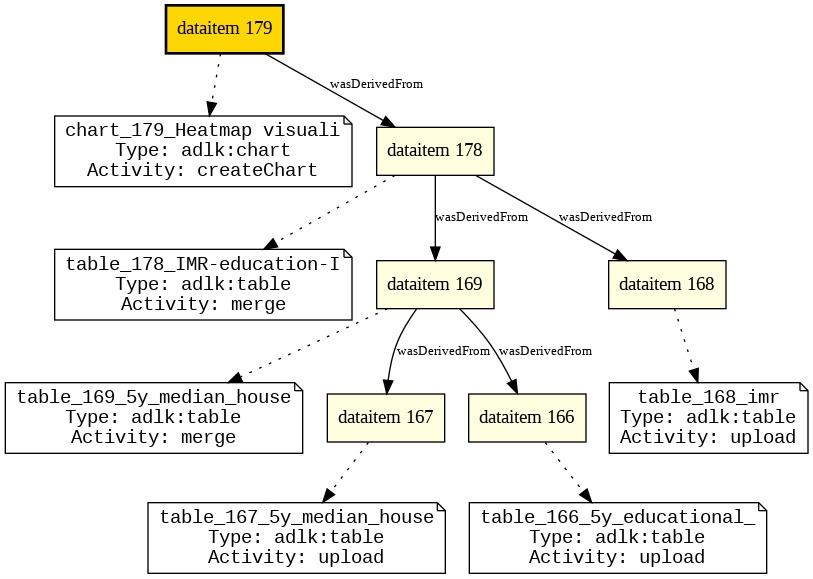}
\caption{\small \textbf{Heatmap Provenance}: `dataitem 179' at the root of the tree depicts the heatmap.}
\label{fig:visualisation-history}
\end{figure}

The final output of the analysis in our example was the heatmap. Now, a user visiting the Datalake would like to understand how exactly the heatmap was created. This becomes as easy as clicking a button on the Datalake where we show the provenance and history of \emph{all} data items along with operations involved in creating the heatmap (Figure \ref{fig:visualisation-history}).


\section{Discussion and Conclusions}
\label{sec:conc}
As is clear from Section \ref{sec:usecase}, our Datalake enables users to perform a number of simple and complex analytical tasks on multiple datasets, while the Datalake infrastructure takes over the task of data management, including provenance and version tracking. The ease of use of the Datalake is key democratizing access to data and good data science practices. Access to data and analysis tools are the most important factor in lowering barriers for NGOs, grassroots organisations and students, who may not be well-versed in using the tools of computer science for data processing.

Our Datalake is a work-in-progress. We will continue to engage with social scientists to discuss their requirements and add features that are essential for research while making these features easy to use. 

\subsubsection*{Acknowledgements}
Funding support by Mphasis AI Lab at Ashoka University.

\bibliography{main}
\bibliographystyle{plain}

\end{document}